**Isolated zero field sub-10 nm skyrmions in ultrathin Co films**

Sebastian Meyer[1,*], Marco Perini[2,*], Stephan von Malottki[1], André Kubetzka[2], Roland Wiesendanger[2], Kirsten von Bergmann[2,#], and Stefan Heinze[1]

[1]Institute of Theoretical Physics and Astrophysics, Christian-Albrechts-Universität zu Kiel, Leibnizstrasse 15, 24098 Kiel, Germany

[2]Department of Physics, University of Hamburg, Jungiusstrasse 11, 20355 Hamburg, Germany

* These authors contributed equally to this work.

# e-mail: kbergman@physnet.uni-hamburg.de

**Due to their exceptional topological and dynamical properties magnetic skyrmions[1] – localized stable spin structures on the nanometre scale – show great promise for future spintronic applications[2-4]. To become technologically competitive, isolated skyrmions with diameters below 10 nm that are stable at zero magnetic field and room temperature are desired[5]. Despite finding skyrmions in a wide spectrum of materials[6-14], the quest for a material with these envisioned properties is still ongoing. Here we report zero field isolated skyrmions with diameters smaller than 5 nm coexisting with 1 nm thin domain walls in Rh/Co atomic bilayers on the Ir(111) surface. These spin structures are characterized by spin-polarized scanning tunnelling microscopy and can also be detected using non-spin-polarized tips due to a large non-collinear magnetoresistance[15]. We demonstrate that sub-10 nm skyrmions are stabilised in these ferromagnetic Co films at zero field due to strong frustration of exchange interaction, together with Dzyaloshinskii-Moriya interaction and a large magnetocrystalline anisotropy.**

The stabilisation of isolated magnetic skyrmions at zero magnetic field in a ferromagnetic material due to the Dzyaloshinskii-Moriya interaction (DMI)[16,17] has been predicted already more than 20 years ago based on a micromagnetic model[18]. After the experimental discovery of magnetic skyrmions[6-9] it has been proposed to use such individual skyrmions in novel storage and logic devices[2-5]. This triggered further theoretical studies with a focus on isolated skyrmions confined in nanostructures[19,20], in ultrathin films[21], and in multilayers[22]. From an experimental point of view ultrathin transition-metal films and transition-metal multilayers have proven to be particularly useful to find novel skyrmion systems since the magnetic interactions can be tuned via interface composition and structure[9,11,13,23,24]. Typically, magnetic skyrmions arise in applied magnetic fields: at room temperature skyrmions with diameters between 30 and 400 nm[10-14] have been realized in multilayers, whereas at cryogenic temperatures isolated skyrmions with diameters down to a few nanometres have been observed[9,25,26]. In zero field, remanent isolated skyrmions were obtained recently in a ferrimagnetic material with diameters down to 16 nm[27], however, in ferromagnets so far they have been realized only with diameters of at least 100 nm in confined structures[12], in multilayers[14] or stabilized by interlayer exchange coupling[28].

Figure 1a and b show spin-resolved differential conductance ($dI/dU$) maps of typical Rh islands on Co/Ir(111) in the magnetic virgin state, measured at different bias voltages. The Co monolayer grows as stripes from the Ir step edges, dominantly pseudomorphic in fcc stacking[29]. The Rh monolayer forms compact islands on Co which are either in hcp stacking (blue areas in Fig. 1b) or in fcc stacking (see Supplementary Fig. S1). On top of the Rh/Co there are a few small second layer Rh islands. We observe small-scale lateral variations in the differential conductance of both the Co and the Rh/Co films, presumably due to some intermixing of the two materials. In the following, we will focus on the magnetic state of the $Rh_{hcp}$/Co bilayer (see Supplementary Fig. S2 for $Rh_{fcc}$/Co).

The $Rh_{hcp}$/Co islands in the $dI/dU$ maps of Fig. 1a dominantly exhibit a two-stage contrast, which we attribute to the spin-polarized contribution to the tunnel current, i.e. the tunnel magnetoresistance (TMR). In combination with experiments using an external magnetic field $B$ (see Fig. 2 and Supplementary Fig. S3) we conclude that the $Rh_{hcp}$/Co is out-of-plane ferromagnetic (FM) and the two-stage contrast represents domains oriented in the two opposite magnetisation directions. The domain walls (DWs) separating oppositely magnetised domains surprisingly do not minimize their length but instead follow meandering paths. This suggests that the position of the DWs is dominated by a combination of small DW energy, i.e. only a small energy penalty for their existence, and an inhomogeneous potential landscape within the film due to some intermixing.

In Fig. 1a the $dI/dU$ signal on the DWs is dominated by the TMR contribution and the signal strength scales with the in-plane magnetisation direction within the wall. This in turn is found to be correlated with the local direction of the wall: all brighter DWs are located at the right side of darker domains in Fig. 1a. Thus the magnetic structures in $Rh_{hcp}$/Co/Ir(111) have a fixed sense of magnetisation rotation. It originates from the DMI, which favours Néel-type DWs and skyrmions with unique rotational sense in thin film systems[9,30,31].

At the bias voltage used in Fig. 1b the $dI/dU$ signal on the DWs is dominated by the tunnel non-collinear magnetoresistance (NCMR) contribution. This contrast is not related to the TMR but arises from changes of the local electronic properties within the non-collinear spin texture[15,32]. For the measurement of Fig. 1b we find a higher NCMR-related $dI/dU$ signal for sample positions where adjacent spins have a larger canting, i.e. bright lines mark the positions of DWs. The strength of this NCMR contribution in first approximation scales with the local mean angle between adjacent magnetic moments[15]. This electronic effect can be exploited to detect DWs and skyrmions also with a non-magnetic metallic electrode and is demonstrated here for the first time in a Co film, a widely used magnetic material.

Small circular domains are found in both of the two oppositely magnetized FM domains of $Rh_{hcp}$/Co/Ir(111), see boxes in Fig. 1a and b. The experimental $dI/dU$ maps can be reproduced by scanning tunnelling microscopy (STM) simulations of skyrmions, see grey scale images in Fig. 1c and d, with different contributions of TMR and NCMR for the two different bias voltages (see Supplementary Text S5 for details). The diameters of the two skyrmions are 4.3 nm and 3.5 nm, and the comparison between experimental and simulated line profiles is reasonable (Fig. 1c and d). In Fig. 1e the derived out-of-plane magnetisation components are plotted, which show the typical continuous magnetisation rotation across a skyrmion. It is quite remarkable, that these opposite magnetic skyrmions appear in the virgin state of our Rh/Co film and that they do not collapse regardless of their small diameter.

Beside the occurrence of isolated skyrmions in the FM virgin state of $Rh_{hcp}$/Co (Fig. 1, Supplementary Fig. S3 and S4), they can be induced by shrinking larger FM domains in opposite magnetic field as seen in the measurements of Fig. 2a and b: the closed loop domain wall imaged bright in the $dI/dU$ map of Fig. 2a due to the NCMR encloses an isolated FM domain in zero field, which shrinks in size upon application of an out-of-plane magnetic field (Fig. 2b). The $dI/dU$ signal within the white rectangles is plotted versus the lateral position across the magnetic objects in Fig. 2d and e. The solid lines show the simulated NCMR signal for two straight 180° domain walls (Fig. 2d) and a cut through a magnetic skyrmion (Fig. 2e) for a DW width $w$ of 0.8 nm; the corresponding spin structures are shown in Fig. 2c, where the atomic magnetic moments are coloured according to their out-of-plane magnetisation components. The experimental data is reproduced well by the simulated NCMR signal and the derived skyrmion diameter is 2.8 nm, see Fig. 2f, which corresponds to about 10 atomic distances between opposite in-plane magnetisations. We would like to emphasise that there is no plateau in the centre of this circular domain, instead the spins rotate continuously.

To understand why in these Rh/Co films small magnetic skyrmions are stable at zero magnetic field, we apply density functional theory (DFT) (see methods). Figure 3a shows the calculated energy dispersion $E(\boldsymbol{q})$ of homogeneous spin spirals in Rh/Co/Ir(111) along the high symmetry directions of the two-dimensional Brillouin zone (2D-BZ) neglecting spin-orbit coupling (SOC); to reveal the role of the Rh overlayer the system of uncovered Co/Ir(111) is also shown. All Co films have a large FM nearest neighbour exchange constant, as evident from the large energy difference of the FM state at $\bar{\Gamma}$ and the antiferromagnetic states at the BZ boundary. At small $|\boldsymbol{q}|$, i.e. for small angles between nearest neighbour moments, Co/Ir(111) shows a rise of the energy with $q^2$, as expected for a typical ferromagnet (Fig. 3b). In contrast, $E(\boldsymbol{q})$ is extremely flat for Rh/Co/Ir(111) for both Rh stackings and to describe the dispersion, a $q^4$ term is required (Fig. 3b). Such an energy dispersion is characteristic for strong exchange frustration, where antiferromagnetic interactions beyond nearest neighbours compete with FM exchange between nearest neighbours (see Supplementary Tables S11 and S12 for values). This is quite unexpected for Co films, but, as we will show, it turns out to be beneficial for the stabilisation of small magnetic skyrmions in zero field.

Calculations including SOC show a strong out-of-plane magnetocrystalline anisotropy energy for both Rh/Co stackings (with 1.2 and 1.6 meV/Co atom, for fcc- and hcp-Rh stacking, respectively). In addition, they reveal an energy contribution to cycloidal spin spirals due to the DMI (Fig. 3c) favouring a clockwise rotational sense. To explore the resulting spin structures we apply atomistic spin dynamics using the parameters from DFT (see methods and Supplementary Tables S11 and S12). For $Rh_{fcc}$/Co we find out-of-plane FM domains and clockwise rotating domain walls with a width of 1.4 nm, see Fig. 4a. The domain walls are exceptionally thin because of the large magnetocrystalline anisotropy in combination with the flat spin spiral dispersion for $|\boldsymbol{q}| < 0.2 \frac{2\pi}{a}$. The latter is responsible for an almost vanishing energy cost for spin cantings between nearest neighbours of up to almost 20° (cf. Fig. 3c). The DW energy amounts to only 2.0 meV/nm, one order of magnitude smaller than for Co/Ir(111) or Pt/Co/Ir(111)[31]. For $Rh_{hcp}$/Co the DW energy is negative due to the spin spiral energy minimum of $E = -0.7$ meV/Co-atom with respect to the FM state. However, the DMI is typically overestimated in our approach[33] and intermixing at the Rh/Co interface, as suggested by the STM measurements, will lead to an additional reduction of the DMI (cf. Ref. [34] and Supplementary Fig. S7). Therefore, we reduce the DMI in the simulations for $Rh_{hcp}$/Co below the critical value to obtain a FM ground state as in the experiment (see Supplementary Table S12 for values). The resulting DW width is 1.3 nm, see Fig. 4a, similar to the experimental value. With this reduced DMI the DW energy is positive, and its small value of only 4.4 meV/nm agrees well with the experimental observation of meandering domain walls with a path dominated by the inhomogeneous potential landscape due to intermixing.

In agreement with the experimental findings our spin dynamics simulations show small zero magnetic field skyrmions within the FM ground state of the Rh/Co films (Fig. 4a). The small skyrmion diameter of about 4 nm is the result of the combination of flat energy dispersion of spin spirals close to the FM state (Fig. 3c) which reduces the energy cost of a fast spin rotation, and large magnetocrystalline anisotropy, which enforces the fast spin rotation. Note that all previously found nanometre-sized isolated skyrmions in ultrathin films[9,25,35] were induced by a magnetic field out of a spin spiral ground state and do not exist in zero magnetic field.

To gain further insight into the mechanism that stabilises these small skyrmions at zero field, we calculate the energy barrier using minimum energy path calculations (see methods). Typically, skyrmion annihilation into the FM state occurs via a collapse in which the skyrmion shrinks in size and the energy barrier stems from DMI (cf. Supplementary Fig. S8). For this radial symmetric collapse mechanism the nearest-neighbour exchange interaction lowers the barrier. However, frustrated exchange interactions can provide a contribution that increases the energy barrier[36]. For $Rh_{hcp}$/Co films the frustration is so strong that the total exchange contribution to the energy barrier even exceeds that of the DMI (Supplementary Fig. S9). To lower the energy barrier, a different annihilation mechanism is preferred at zero magnetic field and Fig. 4b shows how the energy evolves along the minimum energy path starting from the skyrmion and ending in the FM state. At the saddle point (SP) we obtain a large energy barrier of about 300 meV. The contribution due to exchange is more than three times larger than that of the DMI. Decomposing the exchange contribution by the neighbouring shells (see Fig. 4b right) shows that the barrier unexpectedly results from the large FM nearest-neighbour exchange interaction ($J_1$).

One can understand the individual contributions to the energy barrier by looking at the annihilation mechanism (Fig. 4c): just before the SP (SP-1) the spin structure is a slightly oval shaped skyrmion with a similar size as the initial skyrmion. At the SP a singular point is formed in the in-plane magnetized region of the skyrmion. This is energetically very unfavourable in terms of nearest-neighbour FM exchange interaction $J_1$. In contrast, the energy cost due to DMI is much lower since only a small part of the spins are not rotating with the preferred sense. The formation of the singular point transforms the skyrmion into a so-called chimera skyrmion[37] (SP+1) with a vanishing topological charge which easily collapses into the FM state. This novel annihilation mechanism is preferred over the radial symmetric skyrmion collapse at zero field because it greatly reduces the DMI energy barrier and the stability of zero-field skyrmions in this system is a result of the frustrated exchange interaction.

Our work demonstrates that isolated skyrmions with a diameter of below 5 nm can be stabilised without applied magnetic field in ultrathin ferromagnetic Co films due to strong exchange frustration together with moderate DMI and large magnetocrystalline anisotropy. We have shown that the detection of individual sub-10 nm skyrmions via the NCMR effect is possible also in Co films, meaning that these skyrmions can be directly detected in an all-electrical read-out. We anticipate that multilayers can be tailored to transfer these advantageous properties to structures suitable for applications at room temperature.

**Methods:**

**Sample preparation.** The Ir(111) single crystal surface was cleaned by cycles of annealing in oxygen with partial pressures in the range of $10^{-7}$ mbar up to about 1800 K to remove C impurities. For each sample preparation the surface was sputtered with Ar ions of about 800 eV with subsequent annealing to around 1500 K for 60 seconds. The Co was deposited onto the substrate at elevated temperatures to achieve step flow growth. The Rh was deposited after the sample had reached room temperature. Typical deposition rates for Co and Rh are between 0.1 and 0.2 atomic layers per minute. Samples were transferred in vacuo to a low temperature scanning tunnelling microscope equipped with a Cr bulk tip for spin-resolved measurements. The Cr bulk tip was introduced into ultra-high vacuum after etching and in-situ cleaning was performed by field emission.

**Density functional theory.** We apply DFT based on the full potential linearized augmented plane wave (FLAPW) method as implemented in the FLEUR code (www.flapw.de). This all-electron method ranks among the most accurate implementations of DFT. Computational details for Co/Ir(111) can be found in Ref. [32]. For Rh/Co/Ir(111) we performed structural relaxations within the ferromagnetic state using a symmetric film consisting of a Rh/Co bilayer on both sides of five layers of Ir(111) with the theoretical lattice constant ($a = 3.82$ Å)[39]. Relaxed interlayer distances are given in Supplementary Table S13. The muffin tin (MT) radii were $R_{MT} = 2.31$ a.u. for Ir and Rh and $R_{MT} = 2.23$ a.u. for Co. The cutoff for the basis functions was $k_{max} = 4.0$ a.u.$^{-1}$. We used 240 $\boldsymbol{k}$-points in the irreducible wedge of the two-dimensional Brillouin zone (2D-BZ) and the generalized gradient approximation[40].

For spin spiral calculations[41,42] with and without spin-orbit coupling and to obtain the magnetocrystalline anisotropy energy, an asymmetric slab with a Rh/Co bilayer on nine layers of the Ir(111) substrate was used. The energy dispersion of flat spin spirals (see Supplementary Text S14 for details) are calculated with a dense $\boldsymbol{k}$-point mesh of 44 × 44 $\boldsymbol{k}$-points in the full 2D-BZ and a basis cutoff of $k_{max} = 4.0$ a.u.$^{-1}$ was used. Close to the $\bar{\Gamma}$-point ($|\boldsymbol{q}| \to 0$), we checked the convergence of the energy dispersions with up to 100 × 100 $\boldsymbol{k}$-points and $k_{max} = 4.3$ a.u.$^{-1}$. These calculations were performed in local density approximation[43].

**Spin-dynamics simulations.** In order to relax the spin structures of the domain walls and the isolated skyrmions and to calculate their energy differences with respect to the FM state, we used the Landau-Lifshitz equation:

$$\hbar \frac{d\boldsymbol{m}_i}{dt} = \frac{\partial \mathcal{H}}{\partial \boldsymbol{m}_i} \times \boldsymbol{m}_i - \alpha \left( \frac{\partial \mathcal{H}}{\partial \boldsymbol{m}_i} \times \boldsymbol{m}_i \right) \times \boldsymbol{m}_i \qquad (1)$$

where $\boldsymbol{m}_i = \frac{\boldsymbol{M}_i}{M_i}$ is the unit vector of the magnetic moment at atom site $i$, $\alpha$ is the damping parameter and $\mathcal{H}$ is the Hamiltonian:

$$\mathcal{H} = -\sum_{ij} J_{ij}(\boldsymbol{m}_i \cdot \boldsymbol{m}_j) - \sum_{ij} \boldsymbol{D}_{ij}(\boldsymbol{m}_i \times \boldsymbol{m}_j) + K \sum_i (m_i^z)^2 \qquad (2)$$

Here $J_{ij}$ denotes the strength of the exchange interaction between spins on atom sites $i$ and $j$ and $\boldsymbol{D}_{ij}$ is the vector characterizing their DMI. $K$ represents the strength of the uniaxial anisotropy. For the simulations presented in Fig. 4, we have used various damping parameters of $\alpha \in [0.05,1]$ and a time step of 0.1 fs. The values of the exchange constants, the DMI, and the magnetocrystalline anisotropy energy are given in the Supplementary Tables S11 and S12. We used a hexagonal lattice of 70 × 70 spins and a magnetic moment of 2.5 $\mu_B$ which corresponds to a combined value of Rh (0.6 $\mu_B$), Co (1.8 $\mu_B$) and the Ir interface layer (0.1 $\mu_B$) in the FM state.

**Geodesic nudged elastic band method.** Using the relaxed structures of skyrmions from spin-dynamics simulations, we calculate the minimum energy paths (MEPs) for annihilation processes using the geodesic nudged elastic band (GNEB) method[44]. Starting from a local energy minimum (skyrmion), a path is generated into the global energy minimum (FM state). The path is systematically brought into the MEP, while it is divided into a discrete chain of states, the so-called images. The first image corresponds to the skyrmion and the last image to the FM state. After the relaxation of the starting point, the effective field is calculated along a local tangent to the path at each image. Its components are substituted by an artificial spring force between the images to ensure a uniform distribution of the path. Once, the whole chain of images is converged, the path represents the MEP and the saddle point (SP) is the energy maximum. To determine the height and the position of the SP correctly, the climbing image technique is applied.

**Acknowledgements:**

This project has received funding from the European Union's Horizon 2020 research and innovation programme under grant agreement No 665095 (FET-Open project MAGicSky). K.v.B. and A.K acknowledge financial support from the Deutsche Forschungsgemeinschaft (DFG, German Research Foundation) - 402843438; - 408119516. S.M., S.v.M., and S.H. thank the Norddeutscher Verbund für Hoch- und Höchstleistungsrechnen (HLRN) for providing computational resources. We thank Pavel F. Bessarab for valuable discussions.

**Author contribution statement**

S.H. devised the project. M.P. performed the measurements. M.P., K.v.B., and A.K. analysed the experimental data. S.M. performed the DFT calculations. S.M. and S.v.M. performed the spin dynamics and GNEB calculations. S.M., S.v.M., and S.H. analysed the calculations. M.P. and S.M. prepared the figures. M.P., K.v.B., S.M., and S.H. wrote the manuscript. All authors discussed the results and contributed to the manuscript.

**Competing financial interests**

The authors declare no competing financial interests.

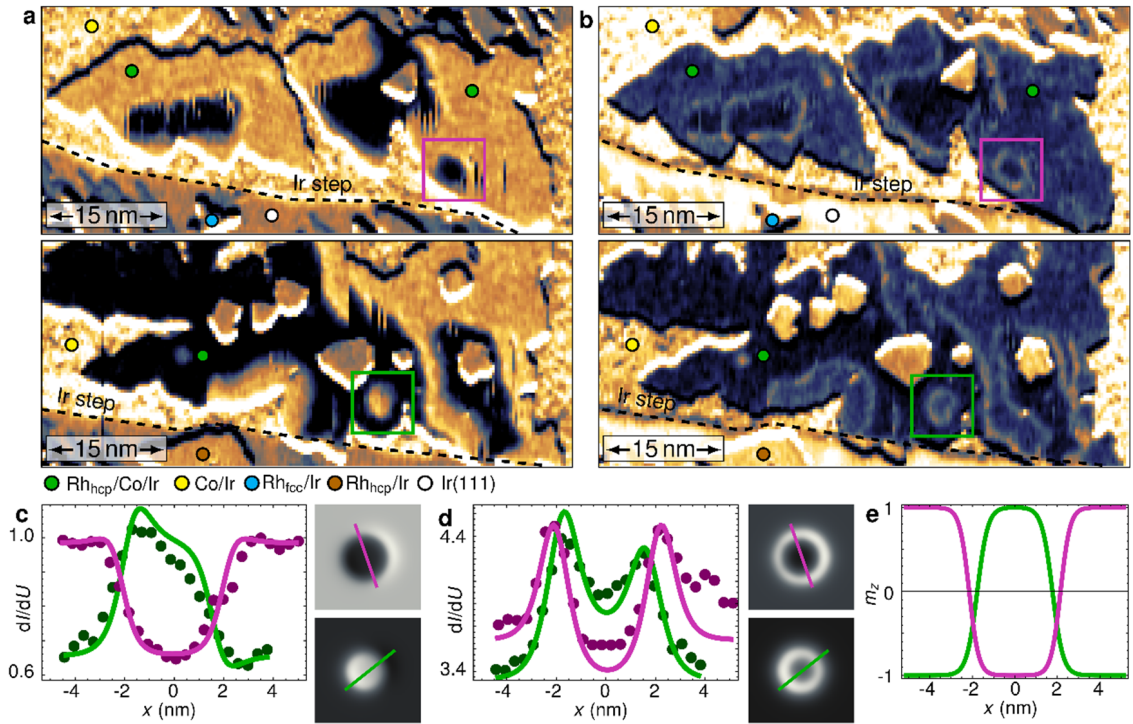

**Figure 1 | Zero field skyrmions in Rh/Co/Ir(111). a,b,** Spin-resolved differential conductance maps of 0.4 atomic layers of hcp-stacked Rh on 0.5 atomic layers of Co on Ir(111) measured at different bias voltages. Whereas in **a** the TMR contribution dominates and the FM domains can be identified by their two-stage contrast, in **b** the NCMR contribution is strong and the domain walls appear as bright lines. The magnetic tip is identical for these measurements and it is sensitive to both the out-of-plane and an in-plane component of the sample's magnetisation. Isolated skyrmions with opposite magnetisation are indicated by the boxes (**a**: $U$ = -250 mV; **b**: $U$ = -400 mV; **a,b**: $I$ = 800 pA, $B$ = 0 T, $T$ = 4.2 K, Cr bulk tip, fast scan axis is vertical). **c,d,** d$I$/d$U$ signal (dots) across the isolated skyrmions in a,b, together with line profiles (solid lines) of STM simulations of two skyrmions (see Supplementary Text S5). **e,** Derived out-of-plane magnetisation component $m_z$ across the two skyrmions.

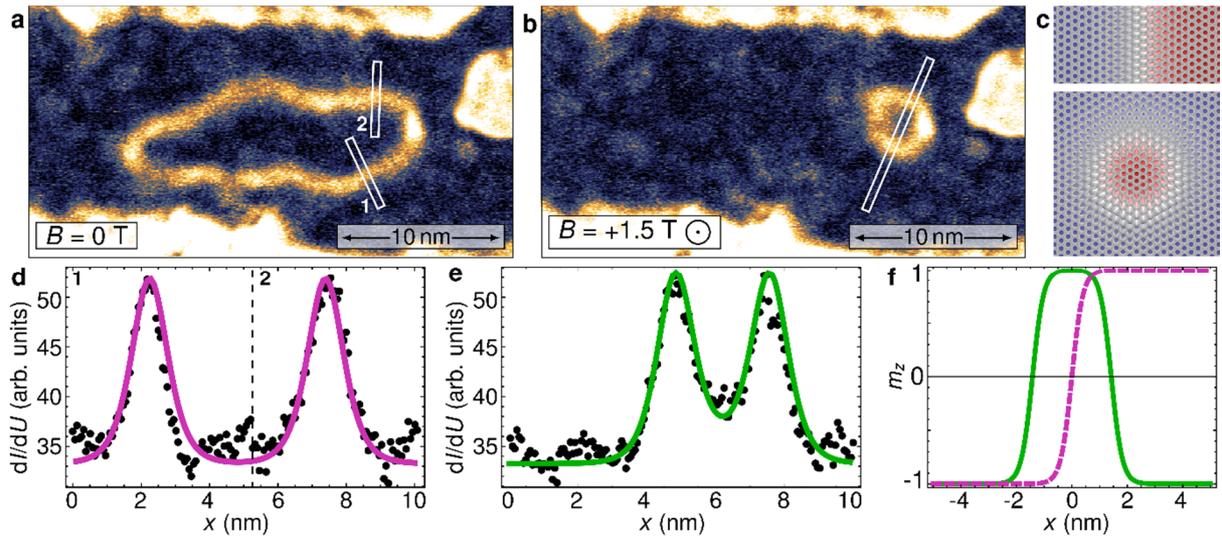

**Figure 2 | Domain wall width and skyrmion diameter. a,b,** d$I$/d$U$ maps of an isolated magnetic object in Rh$_{hcp}$/Co/Ir(111) at zero field and at $B$ = +1.5 T measured with a non-spin-polarized tip; because of the NCMR contribution the domain walls have a higher signal than the FM domains ($U$ = -250 mV, $I$ = 800 pA, $T$ = 4 K, Cr bulk tip). **c,** sketches of a domain wall and a skyrmion ($w$ = 0.8 nm), the colour of the atomic magnetic moments indicates the out-of-plane magnetization component. **d,e,** d$I$/d$U$ signal (dots) across the two domain walls and the skyrmion of **a,b**, positions are indicated by the white rectangles; the solid lines are profiles of STM simulations of the domain wall and skyrmion shown in **c**. **f:** Corresponding out-of-plane magnetisation component $m_z$ across the domain wall and skyrmion of **c-e**.

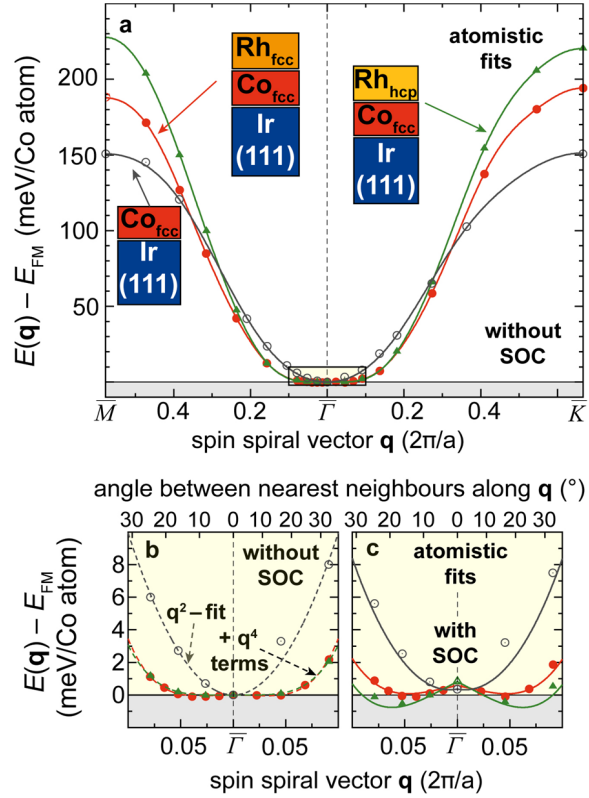

**Figure 3 | Energy dispersion of spin spirals for ultrathin Co films. a,** Energy dispersion of planar homogeneous spin spirals for Co/Ir(111), Rh$_{fcc}$/Co/Ir(111) and Rh$_{hcp}$/Co/Ir(111) along the high symmetry directions $\bar{M} - \bar{\Gamma} - \bar{K}$ of the two-dimensional Brillouin zone without spin-orbit coupling (SOC). Energies are given with respect to the FM state. The symbols (open grey circles for Co/Ir(111), red filled circles for Rh$_{fcc}$/Co/Ir(111) and green triangles for Rh$_{hcp}$/Co/Ir(111)) represent the DFT calculations while the solid lines are the fits to the Heisenberg exchange interaction beyond nearest neighbours (see Supplementary Text S14 for details). **b,** Zoom around the FM area ($\bar{\Gamma}$-point) of **a**, where dashed lines represent a $q^2$ fit in the case of Co/Ir(111) and a fit including an additional $q^4$ term in the case of Rh/Co/Ir(111). **c,** Zoom around the FM area ($\bar{\Gamma}$-point) of the energy dispersion for cycloidal spin spirals including the effect of spin-orbit coupling. The DMI leads to the local energy minima for clockwise rotating spin spirals along both high symmetry directions and the magnetocrystalline anisotropy energy is responsible for the constant energy shift of the spin spirals with respect to the FM state.

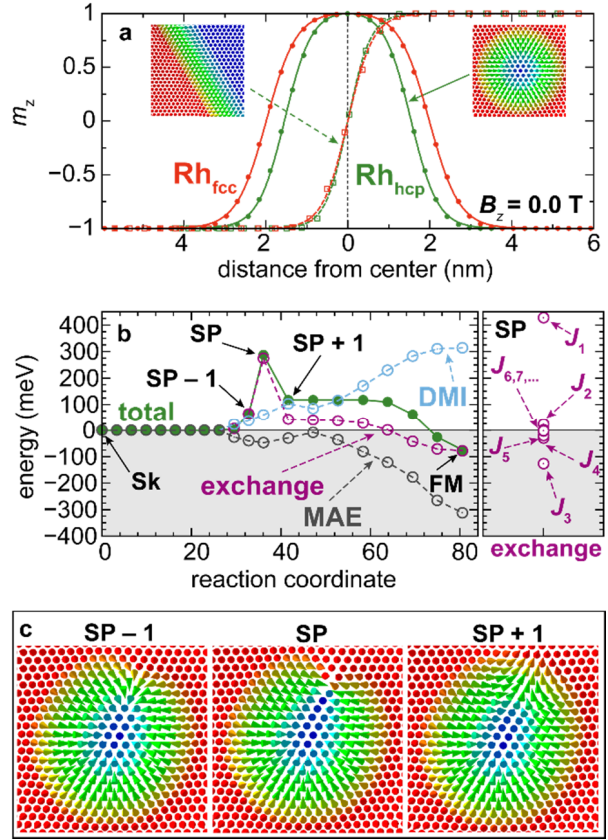

**Figure 4 | Profile and stability of zero-field skyrmions in Rh/Co/Ir(111). a,** Domain wall profiles (open symbols) and zero field skyrmion profiles (filled symbols), i.e. z-component $m_z$ of the local magnetic moment, obtained based on atomistic spin dynamics for $Rh_{hcp}$/Co/Ir(111) (green curves) and $Rh_{fcc}$/Co/Ir(111) (red curves). The dashed and solid lines are fits to the standard domain wall and skyrmion profile[8], respectively. Note that there are small deviations in both cases due to the exchange frustration. The insets show the domain wall and skyrmion spin structures on the two-dimensional atomic lattice. **b,** Total energy and energy contributions from the different interactions (exchange, DMI, magneto-crystalline anisotropy energy MAE) at zero field versus the reaction coordinate along the minimum energy path from the initial isolated skyrmion (Sk) state to the final ferromagnetic (FM) state for $Rh_{hcp}$/Co/Ir(111) (the same annihilation mechanism occurs for fcc Rh stacking). Energies are summed over all atoms of the simulation box and are given relative to the energy of the isolated skyrmion state. The saddle point (SP) is indicated. To the right: exchange energy at the saddle point, $E_{SP}$, resolved with respect to the exchange interactions of different shells: $J_{1..10}$. **c,** Images along the minimum energy path in **b** right before the SP (SP-1), at the SP, and just after the SP (SP-1). Each data point in **b** corresponds to one image. For all images of the minimum energy path see Supplementary Fig. S10. Note that in **c** only a small part of the full simulation box which contained (70×70) spins is shown.